\documentclass[preprint,12pt]{elsarticle}

\usepackage{amssymb}
\usepackage{amsmath}

\begin{document}

\begin{frontmatter}
\title{Cryptography based on operator theory (\uppercase\expandafter{\romannumeral1}): quantum no-key protocols}
\author[label1]{Li Yang}\ead{yangli@iie.ac.cn}
\author[label2]{Min Liang}
\address[label1]{State Key Laboratory of Information Security, Institute of Information Engineering, Chinese Academy of Sciences,
Beijing 100093, China}
\address[label2]{Data Communication Science and Technology Research Institute, Beijing 100191, China}

\begin{abstract}
We study cryptography based on operator theory, and propose quantum no-key (QNK) protocols from the perspective of operator theory, then present a framework of QNK protocols. The framework is expressed in two forms: trace-preserving quantum operators and natural presentations. Then we defined the information-theoretical security of QNK protocols and the security of identification keys. Two kinds of QNK protocols are also proposed. The first scheme is constructed based on unitary transformation, and the other is constructed based on two multiplicative commutative sets.
\end{abstract}


\end{frontmatter}

\makeatletter
    \newcommand{\rmnum}[1]{\romannumeral #1}
    \newcommand{\Rmnum}[1]{\expandafter\@slowromancap\romannumeral #1@}
    \newcommand{\bm}[1]{\mbox{\boldmath{$#1$}}}
\makeatother

\section{Introduction}
The earliest group of quantum message oriented protocols is suggested in \cite{Boykin00, Ambainis00, Nayak07}, which can be regarded as a quantum version of one-time pad, the sender and the receiver must preshare secretly a classical key. Later, a public-key encryption scheme of quantum message is proposed \cite{Yang03}. Recently, this kind of public-key cryptosystems has been developed~\cite{Yang10}.

Here we consider another technique to securely transmit quantum message, so called quantum no-key (QNK) protocol. No-key protocol was first proposed by Shamir \cite{Menezes97}. It is a wonderful idea to transmit classical messages secretly in public channel, independent of the idea of public-key cryptosystem and that of secret-key cryptosystem. However, the protocol presented is computationally secure, cannot resists a man-in-the-middle(MIM) attack. \cite{Yangli02a,Yangli02} develop a quantum from of no-key protocol based on single-photon rotations, which can be used to transmit classical and quantum messages secretly. It can be seen that the security of the QNK protocol is based on the laws of quantum mechanics, so it is beyond computational hypothesis. Ref. \cite{Yangli03} proposed a protocol based on quantum computing of Boolean functions. This protocols is constructed with inherent identifications in order to prevent MIM attack. Similar to the idea of QNK protocol, Kanamori et al.\cite{Kanamori05} proposed a protocol for secure data communication, Kye et al.\cite{Kye05} proposed a quantum key distribution scheme, and Kak \cite{Subhash07} proposed a three-stage quantum cryptographic protocol for key agreement. Wu and Yang \cite{Wu09} presents a practical QNK protocol, and studied a new kind of attack named unbalance-of-information-source (UIS) attack. This kind of attack may also be effective to quantum secure direct communication protocols, such as those in \cite{Beige01,Bostrom02,Deng2003,Deng2004}.

In the paper, the theory of QNK protocols is studied, and a framework of QNK protocols is presented. Then we defined the information-theoretical security of QNK protocols, and the security of identification keys. Finally, two kinds of QNK protocols are presented.

\section{Quantum no-key protocols}
\subsection{Framework}
Let us consider a general framework of quantum no-key protocol, in which two ancillary states are used. Suppose Alice will send quantum message $\rho\in H_M$. The ancillary states used by Alice and Bob are $\rho_A$ and $\rho_B$, respectively. The framework of QNK protocol is described as (see Figure~\ref{fig1}):

\begin{figure}[htp!]
\begin{center}
\includegraphics[width=12cm]{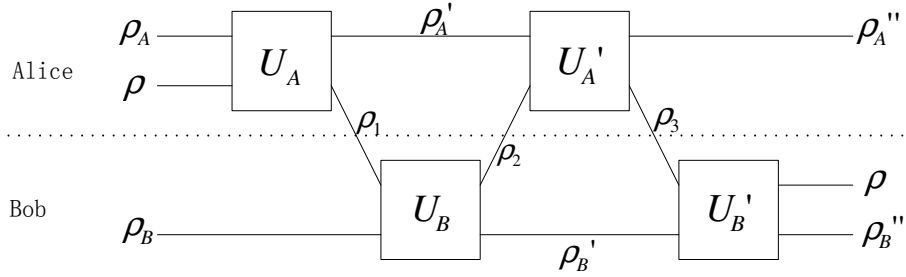}
\end{center}
\caption{\label{fig1} A general framework of quantum no key protocol. This figure is divided into two part by a dashed line. The part above the dashed line describes Alice's operations, and the other part describes Bob's operations. The quantum state $\rho$ is the plain state, and $\rho_1,\rho_2,\rho_3$ represents the three cipher states transmitted between Alice and Bob. $\rho_A,\rho_B$ are two ancillary states generated randomly by Alice and Bob, respectively.}
\end{figure}

\begin{enumerate}
\item{}
Alice randomly prepare a quantum state $\rho_A$, then performs $U_A$ on the quantum states $\rho_A\otimes\rho$ and gets $U_A(\rho_A\otimes\rho)U_A^\dagger$. Then she sends to Bob the first cipher state $\rho_1$, $$\rho_1=tr_A(U_A(\rho_A\otimes\rho)U_A^\dagger)\triangleq \mathcal{E}_A(\rho).$$
She retains the state $\rho_A'=tr_M(U_A(\rho_A\otimes\rho)U_A^\dagger)$.
\item{}
Bob randomly prepares a quantum state $\rho_B$, then performs $U_B$ on the quantum states $\rho_1\otimes\rho_B$  and gets $U_B(\rho_1\otimes\rho_B)U_B^\dagger$. Then he sends to Alice the second cipher state $\rho_2$, $$\rho_2=tr_B(U_B(\rho_1\otimes\rho_B)U_B^\dagger)\triangleq \mathcal{E}_B(\rho_1).$$
He retains the state $\rho_B'=tr_M(U_B(\rho_1\otimes\rho_B)U_B^\dagger)$.
\item{}
Alice performs $U_A'$ on $\rho_A'\otimes\rho_2$, and sends to Bob the third cipher state $\rho_3$, $$\rho_3=tr_A(U_A'(\rho_A'\otimes\rho_2)U_A'^\dagger)\triangleq\mathcal{E}_A'(\rho_2).$$
\item{}
Bob performs $U_B'$ on $\rho_3\otimes\rho_B'$, and gets the message $\rho\prime$, $$\rho'=e^{i\phi}\rho=tr_B(U_B'(\rho_3\otimes\rho_B')U_B'^\dagger)\triangleq\mathcal{E}_B'(\rho_3).$$
\end{enumerate}

In the above protocols, the four quantum operations $\mathcal{E}_A,\mathcal{E}_B,\mathcal{E}_A',\mathcal{E}_B'$ are all trace-preserving quantum operators. This protocol is correct if and only if the four quantum operators satisfy this condition:
\begin{equation}\label{eqn5}
\mathcal{E}_B'\circ\mathcal{E}_A'\circ\mathcal{E}_B\circ\mathcal{E}_A=e^{i\phi}\mathcal{I}.
\end{equation}

{\bf Remark:} As a special case, the unitary transformations $U_A$,$U_B$ can be chosen as bitwise controlled-unitary transformations where the message qubits act as control qubits, and $U_A'=U_A^\dagger$,$U_B'=U_B^\dagger$. In this case, $(I\otimes U_B)(U_A \otimes I)=(U_A \otimes I)(I\otimes U_B)$, and $(I\otimes U_B')(U_A' \otimes I)(I\otimes U_B)(U_A \otimes I)=I$.

\subsection{Natural Representation}
Trace-preserving quantum operator $\mathcal{E}$ can be written as the form of operator-sum representation
\begin{align}
\mathcal{E}(\rho)=\sum_i E_{i}\rho E_{i}^\dagger.
\end{align}
Its natural representation \cite{Watrous08} is
\begin{align}\label{eqn7}
\overrightarrow{\mathcal{E}}(\rho)= \sum_i (E_{i}\otimes E_{i}^*)\overrightarrow{\rho} \triangleq B\overrightarrow{\rho},
\end{align}
where $\overrightarrow{\rho}$ is a the column vector, and represents the vector form of the density matrix $\rho$.

The quantum operations $\mathcal{E}_1$ and $\mathcal{E}_2$ are trace-preserving quantum operator, and their natural representations are denoted as $A,B$ respectively. Because trace-preserving quantum operator can be realized physically, the operators $A,B$ can be realized physically.

Suppose four operators $A,B,A',B'$ represent the natural representation of the four trace-preserving quantum operators $\mathcal{E}_A,\mathcal{E}_B,\mathcal{E}_A',\mathcal{E}_B'$.
Next we deduce the expressions of the natural representation $A,B,A',B'$. Let the ancillary states of Alice and Bob are $\rho_A=\rho_B=|0\rangle\langle 0|$, the orthogonal basis on the ancillary space is the set $\{e_k\}_k$. From the trace-preserving quantum operator $\mathcal{E}_A(\rho)=tr_A(U_A(|0\rangle\langle 0|\otimes\rho)U_A^\dagger)$, it can be inferred that
\begin{eqnarray*}
\mathcal{E}_A(\rho)&=&\sum_k\langle e_k|U_A(|0\rangle\langle 0|\otimes\rho)U_A^\dagger|e_k\rangle \\
&=&\sum_k E_k\rho E_k^\dagger,
\end{eqnarray*}
where $E_k\equiv\langle e_k|U_A|0\rangle$ is an operator acting on the message space $H_M$. Thus, according to Eq.(\ref{eqn7}), one know that the natural representation of trace-preserving quantum operator $\mathcal{E}_A$ is as follows
$$A=\sum_k E_k\otimes E_k^*=\sum_k \langle e_k|U_A|0\rangle\otimes\langle e_k|U_A^*|0\rangle.$$
Similarly, one can present the natural representations of the other three trace-preserving quantum operators, for example $B=\sum_k \langle e_k|U_B|0\rangle\otimes\langle e_k|U_B^*|0\rangle$.

From the above analysis, the four transformations to the quantum message $\rho$ in the quantum no-key protocol can be described in the form of natural representation.
\begin{eqnarray*}
\overrightarrow{\rho}&\longrightarrow&\overrightarrow{\rho_1}=A\overrightarrow{\rho}\\
&\longrightarrow&\overrightarrow{\rho_2}=B\overrightarrow{\rho_1}\\
&\longrightarrow&\overrightarrow{\rho_3}=A'\overrightarrow{\rho_2}\\
&\longrightarrow&\overrightarrow{\rho}'=B'\overrightarrow{\rho_3}
\end{eqnarray*}
Quantum no-key protocol can be described in the form of trace-preserving quantum operators or natural representation, and the two forms are equivalent. In the following sections, we use the natural representation, and the operators $A$ and $B$ are both natural representations of trace-preserving quantum operators.
$\mathcal{E}_1\circ\mathcal{E}_2=\mathcal{E}_2\circ\mathcal{E}_1$ if and only if $AB=BA$.

The additive commutator $A,B$ satisfy $$AB=BA+K;$$
The multiplicative commutator $A,B$ satisfy $$BA=e^{i\lambda}AB, \lambda\neq 0,$$ or be written as $B^{-1}A^{-1}BA=e^{i\lambda}I, \lambda\neq 0$.

When $K=\lambda=0$, the additive commutator equals to multiplicative commutator. The multiplicative commutator $BA=e^{i\lambda}AB$ is just additive commutator when $\lambda=0$; Multiplicative commutator is different from additive commutator in only a global phase $e^{i\lambda}$ when $\lambda \neq 0$, and they have no difference in physical implementation.

{\bf Theorem 1}: The multiplicative commutators $A,B$ are unitary matrices. If $BA=e^{i\lambda}AB$, where $\lambda\neq 0$, then the sum of the eigenvalues of the operators $A,B$ are zero, respectively.

{\bf Proof} It means to prove: if $A=\sum_j e^{i\varphi_j}|a_j\rangle\langle a_j|$, then $\sum_j e^{i\varphi_j}=0$; if $B=\sum_j e^{i\phi_j}|b_j\rangle\langle b_j|$, then $\sum_j e^{i\phi_j}=0$. The proof is as follows.

From $BA=e^{i\phi_{AB}}AB$ where $A,B$ are both unitary, one know that $A=e^{i\phi_{AB}}B^\dagger AB$. Because $A$ is a unitary transformation, it has spectral decomposition $A=\sum_j\lambda_j|j\rangle\langle j|$, and each eigenvalue $\lambda_j$ can be written as $e^{i\phi_j}$. Thus,
$\sum_j\lambda_j|j\rangle\langle j|=\sum_je^{i\phi_{AB}}\lambda_jB^\dagger |j\rangle\langle j|B$. Then
\begin{equation}\label{eqn8}
\lambda_k=\sum_je^{i\phi_{AB}}\lambda_j\langle k|B^\dagger|j\rangle\langle j|B|k\rangle=\sum_je^{i\phi_{AB}}\lambda_j|\langle j|B|k\rangle|^2.
\end{equation}
Let the $j$-th row and $k$-th column of $B$ is $b_{jk}$. The two side of Eq.(\ref{eqn8}) is added with the variable $k$, and obtains $\sum_k\lambda_k=e^{i\phi_{AB}}\sum_j\lambda_j\sum_k|b_{jk}|^2=e^{i\phi_{AB}}\sum_j\lambda_j$.
Thus, $\sum_j\lambda_j(e^{i\phi_{AB}}-1)=0$. If $\phi_{AB}\neq 0$, then $\sum_j\lambda_j=0$, that means the sum of the eigenvalues of $A$ is 0. Similarly, one can prove that the sum of the eigenvalues of $B$ is 0.$~\hfill{}\Box$

In the next part of this paper, we consider the case that $A'=A^{-1},B'=B^{-1}$.
In this case, Eq.(\ref{eqn5}) can also be expressed as $B^{-1}A^{-1}BA=e^{i\phi}I$. It can be seen that $A,B$ are multiplicative commutator.

In order to identify personal identification in the protocols, Alice and Bob must preshare a secret key $k,i$, and the multiplicative commutator $A,B$ should satisfy:
\begin{equation}
B^{-1}A^{-1}BA=e^{i\phi}N_k(i).
\end{equation}
Alice and Bob preshare secret identification key $k,i$, where $k\in\mathcal{K}$. From the value of $k$, a operator $N_k$ can be obtained and a set $I(k)$ can be constructed, and the secret key $i\in I(k)$. From $k,i$, we can get a set $L(k,i)$ and a set of operators satisfying
\begin{equation}
S_k(i)=\{A_l^{(i)}(k)|l\in L(k,i)\}\cup \{B_l^{(i)}(k)|l\in L(k,i)\}.
\end{equation}
The set of operators $S_k(i)$ should satisfy the condition: for any two elements $A_{l_1}^{(i)}(k),B_{l_2}^{(i)}(k)\in S_k(i)$, it holds that
\begin{equation}\label{eqn1}
[B_{l_2}^{(i)}(k)]^{-1}[A_{l_1}^{(i)}(k)]^{-1} B_{l_2}^{(i)}(k) A_{l_1}^{(i)}(k) =e^{i\phi(l_1,l_2)}N_k(i).
\end{equation}

Alice and Bob communicates according to $k,i$ and the set $S_k(i)$. The process is as follows
\begin{enumerate}
\item{}
Alice randomly selects $l_1\in L(k,i)$, and performs an operator $A_{l_1}^{(i)}(k)$ on quantum message $\overrightarrow{\rho}$, then obtains a state $\overrightarrow{\rho_1}$. She sends it to Bob.
\item{}
Bob randomly selects $l_2\in L(k,i)$, and performs an operator $B_{l_2}^{(i)}(k)$ on quantum state $\overrightarrow{\rho_1}$, then obtains a state $\overrightarrow{\rho_2}$. He sends it to Alice.
\item{}
According to $l_1$, Alice performs an operator $[A_{l_1}^{(i)}(k)]^{-1}$ on quantum state $\overrightarrow{\rho_2}$, then obtains a state $\overrightarrow{\rho_3}$. She sends it to Bob.
\item{}
According to $l_2$, Bob performs an operator $[B_{l_2}^{(i)}(k)]^{-1}$ on quantum state $\overrightarrow{\rho_3}$, then obtains a state $\overrightarrow{\rho_4}$.
\end{enumerate}

From the condition satisfied by $S_k(i)$, it is deduced that
\begin{equation}
\overrightarrow{\rho_4}=[B_{l_2}^{(i)}(k)]^{-1} [A_{l_1}^{(i)}(k)]^{-1} B_{l_2}^{(i)}(k) A_{l_1}^{(i)}(k) \overrightarrow{\rho} = e^{i\phi(l_1,l_2)}N_k(i) \overrightarrow{\rho}.
\end{equation}
Thus, the quantum state obtained by Bob in the end is $N_k(i) \overrightarrow{\rho}$. He can recovery the quantum message $\overrightarrow{\rho}$ by performing the inverse transformation of $N_k(i)$.

Through the value of $l_1$ is unknown by Bob, Bob randomly selects a value of $l_1'\in L(k,i)$, then it can satisfy the relation in Eq.(\ref{eqn1})(only a little difference on total phase). Thus, the inverse transformation of $N_k(i)$ can also be replace by
$$N_k^{-1}(i)=[A_{l_1'}^{(i)}(k)]^{-1}[B_{l_2}^{(i)}(k)]^{-1}A_{l_1'}^{(i)}(k)B_{l_2}^{(i)}(k), \forall l_1'\in L(k,i).$$ Because the operators $A_l^{(i)}(k)$, $B_l^{(i)}(k)$ are all trace-preserving quantum operators and can be implemented physically, the quantum operator $N_k^{-1}$ is physically implementable. Thus, Bob can recovery the quantum message $\overrightarrow{\rho}$ by performing $N_k^{-1}(i)$ on the state $\overrightarrow{\rho_4}$.

In order to enhance the security of this protocol, we can select a group of similarity transformations $\{T_i|i\in L(k,i)\}$. The set $S_k(i)$ in the above protocol is replaced with another set
\begin{equation}
\tilde{S}_k(i)=\{T_iA_l^{(i)}(k)T_i^{-1}|l\in L(k,i)\}\cup \{T_iB_l^{(i)}(k)T_i^{-1}|l\in L(k,i)\}.
\end{equation}
Similarly, $N_k(i)$ is replaced with $T_iN_k(i)T_i^{-1}$, but it is still denoted as $N_k(i)$ for convenient.
Then according to Eq.(\ref{eqn1}), we know $$T_i[B_{l_2}^{(i)}(k)]^{-1}[A_{l_1}^{(i)}(k)]^{-1} B_{l_2}^{(i)}(k) A_{l_1}^{(i)}(k)T_i^{-1}=e^{i\phi(l_1,l_2)}N_k(i).$$
From the following relation
\begin{align}
& T_i[B_{l_2}^{(i)}(k)]^{-1}[A_{l_1}^{(i)}(k)]^{-1} B_{l_2}^{(i)}(k) A_{l_1}^{(i)}(k)T_i^{-1} \nonumber\\
=& [T_iB_{l_2}^{(i)}(k)T_i^{-1}]^{-1}[T_iA_{l_1}^{(i)}(k)T_i^{-1}]^{-1} T_iB_{l_2}^{(i)}(k)T_i^{-1} T_iA_{l_1}^{(i)}(k)T_i^{-1},
\end{align}
it is inferred that: any two elements in $\tilde{S}_k(i)$ such as $T_iA_{l_1}^{(i)}(k)T_i^{-1},T_iB_{l_2}^{(i)}(k)T_i^{-1}$ satisfy the following relation
\begin{equation}
[T_iB_{l_2}^{(i)}(k)T_i^{-1}]^{-1}[T_iA_{l_1}^{(i)}(k)T_i^{-1}]^{-1} T_iB_{l_2}^{(i)}(k)T_i^{-1} T_iA_{l_1}^{(i)}(k)T_i^{-1}=e^{i\phi(l_1,l_2)}N_k(i).
\end{equation}

\subsection{Security}
In QNK protocol, there are 3 times of transmitting quantum ciphers through public quantum channel. Denote $\rho_i$ as the $i$-th quantum cipher with respect to the attackers, where $i=1,2,3$.

{\bf Definition 1:} QNK protocol is information-theoretically secure, if the three quantum ciphers $\rho_1,\rho_2,\rho_3$ satisfy the following condition: for any positive polynomial $p(.)$, and all sufficiently big number $n$, it holds that
\begin{equation}
D\left(\rho_i,\rho_j\right)<\frac{1}{p(n)},\forall i,j\in\{1,2,3\},
\end{equation}
where $\rho_1,\rho_2,\rho_3$ are all $n$-qubit ciphers.

In fact, this definition equals to the following definition.

{\bf Definition 2:} QNK protocol is information-theoretically secure, if the three quantum ciphers $\rho_1,\rho_2,\rho_3$ satisfy the following condition: there exists a quantum state $\tau$, such that for any positive polynomial $p(.)$ and all sufficiently big number $n$, it holds that
\begin{equation}
D\left(\rho_i,\tau\right)<\frac{1}{p(n)},\forall i\in\{1,2,3\},
\end{equation}
where $\rho_1,\rho_2,\rho_3$ are all $n$-qubit ciphers.

In one hand, if QNK protocol satisfies Definition 2, it can be deduced that $D\left(\rho_i,\rho_j\right)<D\left(\rho_i,\tau\right)+D\left(\rho_j,\tau\right)<\frac{2}{p(n)},\forall i,j\in\{1,2,3\}$, then the protocol satisfies Definition 1;
In the other hand, if QNK protocol satisfies Definition 1, and let $\tau=\rho_1$, we know that $D\left(\rho_i,\tau\right)=D\left(\rho_i,\rho_1\right)<\frac{1}{p(n)},\forall i\in\{1,2,3\}$, then the protocol satisfies Definition 2. Thus, the two definitions are equivalent to each other.

Here, $\rho_i$ is the $i$-th quantum cipher with respect to the attackers, where $i=1,2,3$. In the QNK protocol, $\overrightarrow{\rho_1}$ is obtained by performing quantum transformation $A_{l_1}^{(i)}(k)$ on quantum state $\overrightarrow{\rho}$. However, with regard to the attackers, the random number $l_1$ and authentication key $i,k$ used by Alice cannot be obtained, so
$$\overrightarrow{\rho_1}=\sum_{i,k,l_1}p_i p_k p_{l_1} A_{l_1}^{(i)}(k)\overrightarrow{\rho},$$
where $p_i, p_k$ are the probability of selecting the authentication key $i,k$, and $p_{l_1}$ is the probability of the local number $l_1$ being selected by Alice.
Similarly, the attackers cannot obtain the random number $l_2$ and authentication key $i,k$ used by Bob, so
$$\overrightarrow{\rho_2}=\sum_{i,k,l_1,l_2}p_i p_k p_{l_1} p_{l_2}B_{l_2}^{(i)}(k)A_{l_1}^{(i)}(k)\overrightarrow{\rho},$$
where $p_{l_2}$ is the probability of the local number $l_2$ being selected by Bob.
$$\overrightarrow{\rho_3}=\sum_{i,k,l_1,l_2}p_i p_k p_{l_1} p_{l_2}(A_{l_1}^{(i)}(k))^{-1}B_{l_2}^{(i)}(k)A_{l_1}^{(i)}(k)\overrightarrow{\rho}.$$

%

{\bf Definition 3:} QNK protocol is information-theoretically secure, if the three operators $\sum_{i,k,l_1}p_i p_k p_{l_1}A_{l_1}^{(i)}(k)$,  $\sum_{i,k,l_1,l_2}p_i p_k p_{l_1} p_{l_2}B_{l_2}^{(i)}(k)A_{l_1}^{(i)}(k)$, and \\ $\sum_{i,k,l_1,l_2}p_i p_k p_{l_1} p_{l_2}(A_{l_1}^{(i)}(k))^{-1}B_{l_2}^{(i)}(k)A_{l_1}^{(i)}(k)$ satisfy the condition: for any positive polynomial $p(.)$, and all sufficiently large number $n$, it holds that
\begin{eqnarray*}
&&||\sum_{i,k,l_1}p_i p_k p_{l_1} A_{l_1}^{(i)}(k)-\sum_{i,k,l_1,l_2}p_i p_k p_{l_1} p_{l_2}B_{l_2}^{(i)}(k)A_{l_1}^{(i)}(k)||_{\diamondsuit}<\frac{1}{p(n)},\\
&&||\sum_{i,k,l_1,l_2}p_i p_k p_{l_1} p_{l_2}\left(B_{l_2}^{(i)}(k)A_{l_1}^{(i)}(k)-(A_{l_1}^{(i)}(k))^{-1}B_{l_2}^{(i)}(k)A_{l_1}^{(i)}(k)\right)||_{\diamondsuit}<\frac{1}{p(n)},\\
&&||\sum_{i,k,l_1}p_i p_k p_{l_1}A_{l_1}^{(i)}(k)-\sum_{i,k,l_1,l_2}p_i p_k p_{l_1} p_{l_2}(A_{l_1}^{(i)}(k))^{-1}B_{l_2}^{(i)}(k)A_{l_1}^{(i)}(k)||_{\diamondsuit}<\frac{1}{p(n)}.
\end{eqnarray*}
where the notation $||*||_{\diamondsuit}$ represents diamond norm.

It is obvious that a sufficient condition for information-theoretical security  is as follow:
\begin{eqnarray*}
&&||I-\sum_{l_2}p_{l_2}B_{l_2}^{(i)}(k)||_{\diamondsuit}<\frac{1}{p(n)},\forall i,k,\\
&&||I-A_{l_1}^{(i)}(k)^{-1}||_{\diamondsuit}<\frac{1}{p(n)}, \forall i,k,l_1, \\
&&||I-\sum_{l_2}p_{l_2}(A_{l_1}^{(i)}(k))^{-1}B_{l_2}^{(i)}(k)||_{\diamondsuit}<\frac{1}{p(n)}, \forall i,k,l_1.
\end{eqnarray*}

For the QNK protocol which uses authentication key, when considering its security,  besides analyzing the security of quantum message, we also should analyze the security of the authentication key. Here we present a definition of the security of authentication key in QNK protocol.

{\bf Definition 4:} The authentication key in QNK protocol is secure, if for any positive polynomial $p(.)$, and all sufficiently large number $n$, it holds that
\begin{eqnarray*}
&&||\sum_{l_1}A_{l_1}^{(i_1)}(k_1)-\sum_{l_1}A_{l_1}^{(i_2)}(k_2)||_{\diamondsuit}<\frac{1}{p(n)}, \forall i_1,i_2,k_1,k_2.\\
&&||\sum_{l_1,l_2}B_{l_2}^{(i_1)}(k_1)A_{l_1}^{(i_1)}(k_1)-\sum_{l_1,l_2}B_{l_2}^{(i_2)}(k_2)A_{l_1}^{(i_2)}(k_2)||_{\diamondsuit}<\frac{1}{p(n)}, \forall i_1,i_2,k_1,k_2.\\
&&||\sum_{l_1,l_2}(A_{l_1}^{(i_1)}(k_1))^{-1}B_{l_2}^{(i_1)}(k_1)A_{l_1}^{(i_1)}(k_1)-\sum_{l_1,l_2}(A_{l_1}^{(i_2)}(k_2))^{-1}B_{l_2}^{(i_2)}(k_2)A_{l_1}^{(i_2)}(k_2)||_{\diamondsuit}\\
&&<\frac{1}{p(n)}, \forall i_1,i_2,k_1,k_2.
\end{eqnarray*}

\section{Two schemes of quantum no-key protocols}
\subsection{First scheme}
Unitary transformation is a special kind of trace-preserving quantum operator. Here we assume the quantum operators used by Alice and Bob in QNK protocols are all unitary. Let $N_k(i)=I$. The sets of operators $\{A_l|l\in\{1,2,\ldots,n_A\}\}$ and $\{B_k|k\in\{1,2,\ldots,n_B\}\}$ are natural representations of unitary operator used by Alice and Bob. The two set satisfy the relation:
\begin{equation}\label{eqn4}
B_{l_2}^{-1}A_{l_1}^{-1}B_{l_2}A_{l_1}=e^{i\phi(l_1,l_2)}I.
\end{equation}
The above formula can also be written as $B_{l_2}A_{l_1}=e^{i\phi(l_1,l_2)}A_{l_1}B_{l_2}$.

$A_l,B_k$ are natural representations of unitary operators. So we can assume $A_l=E_l\otimes E_l^*,B_k=F_k\otimes F_k^*$, where $E_l, F_k$ are unitary transformations, and $l\in\{1,2,\ldots,n_A\}$, $k\in\{1,2,\ldots,n_B\}$. Then
\begin{eqnarray}
B_{l_2}^{-1}A_{l_1}^{-1}B_{l_2}A_{l_1}&=&(F_{l_2}\otimes F_{l_2}^*)^{-1}(E_{l_1}\otimes E_{l_1}^*)^{-1}(F_{l_2}\otimes F_{l_2}^*)(E_{l_1}\otimes E_{l_1}^*)\nonumber\\
&=& (F_{l_2}^{-1}E_{l_1}^{-1}F_{l_2}E_{l_1})\otimes(F_{l_2}^{-1}E_{l_1}^{-1}F_{l_2}E_{l_1})^*\nonumber\\
&\triangleq&V_{l_1,l_2}\otimes V_{l_1,l_2}^*,
\end{eqnarray}
where $V_{l_1,l_2}=F_{l_2}^{-1}E_{l_1}^{-1}F_{l_2}E_{l_1}$ is unitary transformation. According to Eq.(\ref{eqn4}), we have $V_{l_1,l_2}\otimes V_{l_1,l_2}^*=e^{i\phi(l_1,l_2)}I_{2^{2n}}$.
From the identity $I_{2^{2n}}=I_{2^n}\otimes I_{2^n}$, it has $V_{l_1,l_2}\otimes V_{l_1,l_2}^*=aI_{2^n}\otimes bI_{2^n}$, where $ab=e^{i\phi(l_1,l_2)}$, and $b=a^*$. So $e^{i\phi(l_1,l_2)}=aa^*=|a|^2$, then $e^{i\phi(l_1,l_2)}=1$. That means it is impossible to produce a global phase in Eq.(\ref{eqn4}). Thus the Eq.(\ref{eqn4}) is rewritten as follows
\begin{equation}\label{eqn6}
B_{l_2}^{-1}A_{l_1}^{-1}B_{l_2}A_{l_1}=I.
\end{equation}

Based on this relation, we construct a QNK scheme, which is described as follows.

Firstly, two sets of operators are selected, such as $$S(A)=\{A_l|l\in\{1,2,\ldots,n_A\}\},$$ $$S(B)=\{B_k|k\in\{1,2,\ldots,n_B\}\}.$$

By using the two sets, Alice and Bob communicate following this QNK protocol.

\begin{enumerate}
\item{}
Alice randomly selects a number $l_1\in\{1,2,\ldots,n_A\}$, then performs quantum operator $A_{l_1}$ on quantum message $\overrightarrow{\rho}$, and obtains the state $\overrightarrow{\rho_1}$. She sends it to Bob.
\item{}
Bob randomly selects a number $l_2\in\{1,2,\ldots,n_B\}$, then performs quantum operator $B_{l_2}$ on quantum message $\overrightarrow{\rho_1}$, and obtains the state $\overrightarrow{\rho_2}$. He sends it to Alice.
\item{}
According to the value of $l_1$, Alice performs quantum operator $A_{l_1}^{-1}$ (or $A_{l_1}^\dagger$) on quantum message $\overrightarrow{\rho_2}$, and obtains the state $\overrightarrow{\rho_3}$. She sends it to Bob.
\item{}
According to the value of $l_2$, Bob performs quantum operator $B_{l_2}^{-1}$ (or $B_{l_2}^\dagger$) on quantum message $\overrightarrow{\rho_3}$, and obtains the state $\overrightarrow{\rho_4}$.
\end{enumerate}

According to the relation (Eq.(\ref{eqn6})) of the two sets $S(A),S(B)$, we know that
$$\overrightarrow{\rho_4}=B_{l_2}^{-1} A_{l_1}^{-1} B_{l_2} A_{l_1} \overrightarrow{\rho} = \overrightarrow{\rho}.$$
Thus, the quantum state obtained by Bob's performing quantum transformation $B_{l_2}^{-1}$ is just the quantum message sent by Alice.

\subsection{Second scheme}
Denote the operation of two operators $A,B$: $(A,B)=B^{-1}A^{-1}BA$.

Suppose there exists two groups of operators $S(A)=\{A_i|i=1,\cdots,n_A\}$ and $S(B)=\{B_i|i=1,\cdots,n_B\}$, which satisfy the following condition
\begin{equation}\label{eqn2}
(A_i,B_j)=N, \forall i\in\{1,\cdots,n_A\},j\in\{1,\cdots,n_B\},
\end{equation}
where $N$ is an operator that is independent of $i,j$.

{\bf Proposition 1:} Suppose two sets of operators $S(A),S(B)$ satisfy the condition Eq.(\ref{eqn2}), then the following relations hold: $\forall i,j$
\begin{eqnarray}
&& (A_i^{-1},B_jB_i^{-1})=I,\\
&& (A_j^{-1},B_jB_i^{-1})=I,\\
&& (B_i,A_jA_i^{-1})=I,\\
&& (B_j,A_jA_i^{-1})=I.
\end{eqnarray}
{\bf Proof:} See the appendix.$\hfill{}\Box$

{\bf Proposition 2:} Suppose two sets of operators $S(A),S(B)$ satisfy the condition Eq.(\ref{eqn2}),  then the following relations hold: $\forall i,j,k,l$
\begin{equation}
(A_iA_j,B_kB_l)=(A_j^2,B_l^2).
\end{equation}
{\bf Proof:} See the appendix.$\hfill{}\Box$

According to Proposition 2, we know the relation $(A_{l_1}A_{k_1},B_{l_2}B_{k_2})=(A_{k_1}^2,B_{k_2}^2)$. Based on this relation, a QNK scheme is constructed as follows.

Suppose the set of similarity transformations $\{T_i|i\in I\}$ is selected. Firstly we construct a set of operators $S_k(i)$ as follows:
\begin{eqnarray}
S_{k_1||k_2}(i)&=&\{T_iA_lA_{k_1}T_i^{-1}|l\in\{1,\cdots,n_A\}\} \nonumber\\
&&\cup\{T_iB_lB_{k_2}T_i^{-1}|l\in\{1,\cdots,n_B\}\}.
\end{eqnarray}

In the set $S_k(i)$, any two elements $A_{l_1}^{(i)}(k_1||k_2)=T_iA_{l_1}A_{k_1}T_i^{-1}$ and $B_{l_2}^{(i)}(k_1||k_2)=T_iB_{l_2}B_{k_2}T_i^{-1}$ satisfy the following relation
\begin{eqnarray}
(A_{l_1}^{(i)}(k_1||k_2),B_{l_2}^{(i)}(k_1||k_2))&=& (T_iA_{l_1}A_{k_1}T_i^{-1},T_iB_{l_2}B_{k_2}T_i^{-1}) \nonumber\\
&=& T_i(A_{l_1}A_{k_1},B_{l_2}B_{k_2})T_i^{-1} \nonumber\\
&=& T_i(A_{k_1}^2,B_{k_2}^2)T_i^{-1}.
\end{eqnarray}
Denote $N_{k_1||k_2}=(A_{k_1}^2,B_{k_2}^2)$, $N_{k_1||k_2}(i)=T_i N_{k_1||k_2} T_i^{-1}$, then
\begin{equation}
(A_{l_1}^{(i)}(k_1||k_2),B_{l_2}^{(i)}(k_1||k_2))=N_{k_1||k_2}(i), ~\forall l_1,l_2.
\end{equation}

Alice and Bob communicate according to $k_1||k_2,i$ ($k_1\in\{1,\cdots,n_A\},k_2\in\{1,\cdots,n_B\}$, $i\in I$) and the set of operators $S_{k_1||k_2}(i)$. The process is as follows.
\begin{enumerate}
\item{}
Alice randomly selects a number $l_1\in\{1,\cdots,n_A\}$, and performs quantum transformation $A_{l_1}^{(i)}(k_1||k_2)$ on quantum message $\overrightarrow{\rho}$, then obtains $\overrightarrow{\rho_1}$. She sends it Bob.
\item{}
Bob randomly selects a number $l_2\in \{1,\cdots,n_B\}$, and performs quantum transformation $B_{l_2}^{(i)}(k_1||k_2)$ on quantum state $\overrightarrow{\rho_1}$, then obtains $\overrightarrow{\rho_2}$. He sends it to Alice.
\item{}
According to $l_1$, Alice performs quantum transformation $[A_{l_1}^{(i)}(k_1||k_2)]^{-1}$ on quantum state $\overrightarrow{\rho_2}$, and obtains $\overrightarrow{\rho_3}$. She sends it to Bob.
\item{}
According to $l_2$, Bob performs quantum transformation $[B_{l_2}^{(i)}(k_1||k_2)]^{-1}$ on quantum state $\overrightarrow{\rho_3}$, and obtains $\overrightarrow{\rho_4}$.
\end{enumerate}

From the condition satisfied by $S_{k_1||k_2}(i)$, it can be deduced that
\begin{equation}
\overrightarrow{\rho_4}=N_{k_1||k_2}(i)\overrightarrow{\rho}=T_i (A_{k_1}^2,B_{k_2}^2) T_i^{-1}\overrightarrow{\rho}.
\end{equation}
Bob performs the inverse transformation of $N_{k_1||k_2}(i)$, which is $[N_{k_1||k_2}(i)]^{-1}=T_i (B_{k_2}^2,A_{k_1}^2) T_i^{-1}$, then the quantum message $\overrightarrow{\rho}$ is recovered.

{\bf Proposition 3:} $(A_{j_1}A_{j_2}A_{j_3},B_{i_1}B_{i_2}B_{i_3})=(A_{j_3}A_{j_2}A_{j_3},B_{i_3}B_{i_2}B_{i_3})$, $\forall i_1,i_2,i_3$, $\forall j_1,j_2,j_3$.

{\bf Proof:} See the appendix.$\hfill{}\Box$

According to Proposition 3, we know the relation $(A_{l_1}A_{k_1}A_{k_2},B_{l_2}B_{k_3}B_{k_4})=(A_{k_2}A_{k_1}A_{k_2},B_{k_4}B_{k_3}B_{k_4})$.
Suppose the set of similarity transformations $\{T_i|i\in I\}$ is selected. We construct the set of operators $S_k(i)$ as follows:
\begin{eqnarray}
S_{k_1||k_2||k_3||k_4}(i)&=&\{T_iA_lA_{k_1}A_{k_2}T_i^{-1}|l=1,\cdots,n_A\} \nonumber\\
&&\cup\{T_iB_lB_{k_3}B_{k_4}T_i^{-1}|l=1,\cdots,n_B\}.
\end{eqnarray}
When the set of operators $S_{k_1||k_2||k_3||k_4}(i)$ is used in the QNK communication, $l$ is a random number selected locally, and $k_1,k_2,k_3,k_4,i$ are authentication keys preshared by Alice and Bob. If Alice selects a local random number $l_1$, then she should perform quantum transformation $T_iA_{l_1}A_{k_1}A_{k_2}T_i^{-1}$; if Bob selects a local random number $l_2$, then he should perform quantum transformation $T_iB_{l_2}B_{k_3}B_{k_4}T_i^{-1}$.
The detailed process of the three interactive communication is the same as the last scheme. After the fourth step, Bob obtains a quantum state
\begin{equation}
\overrightarrow{\rho_4}=T_i (A_{k_2}A_{k_1}A_{k_2},B_{k_4}B_{k_3}B_{k_4}) T_i^{-1}\overrightarrow{\rho}.
\end{equation}
Then Bob performs quantum transformation $T_i (B_{k_4}B_{k_3}B_{k_4},A_{k_2}A_{k_1}A_{k_2}) T_i^{-1}$ and obtains the quantum message $\overrightarrow{\rho}$ sent by Alice.

In general, we have the following relation
\begin{equation}
(A_{l_1}A_{k_1}A_{k_2}\cdots A_{k_n},B_{l_2}B_{k_1'}B_{k_2'}\cdots B_{k_n'})=(A_{k_n}A_{k_1}A_{k_2}\cdots A_{k_n},B_{k_n'}B_{k_1'}B_{k_2'}\cdots B_{k_n'}).
\end{equation}
According to this relation, a new QNK scheme can be constructed similarly. In the new scheme, more operators can be used by Alice and Bob.
\subsection{Construction of the sets $S(A)$ and $S(B)$}
Suppose there exists two groups of operators such as $S(A)=\{A_i|i=1,\cdots,n_A\}$ and $S(B)=\{B_i|i=1,\cdots,n_B\}$, and they satisfy the relation expressed in Eq.(\ref{eqn2}), then $A_i^{-1}B_jA_i=B_jN,\forall j$. Thus $A_{i_1}^{-1}B_jA_{i_1}=A_{i_2}^{-1}B_jA_{i_2}$. That is
\begin{equation}\label{eqn3}
(A_{i_2}A_{i_1}^{-1})B_j=B_j(A_{i_2}A_{i_1}^{-1}),
\end{equation}
or written as $(A_{i_2}A_{i_1}^{-1},B_j)=I$ or $[A_{i_2}A_{i_1}^{-1},B_j]=0$. (Denote $[A,B]=AB-BA$)

According to Eq.(\ref{eqn2}) and the similar deduction, it can be inferred that $\forall i,j,k$,
\begin{eqnarray}
&& [A_iA_j^{-1},B_k]=0, \\
&& [A_iA_j^{-1},B_k^{-1}]=0, \\
&& [B_iB_j^{-1},A_k]=0, \\
&& [B_iB_j^{-1},A_k^{-1}]=0,
\end{eqnarray}

From $[A_iA_j^{-1},B_k]=0,\forall i,j,k$, it can be deduced that
\begin{equation}\label{eqn3}
[A_iA_j^{-1},B_kB_l^{-1}]=[A_iA_j^{-1},B_k]B_l^{-1}+B_k[A_iA_j^{-1},B_l^{-1}]=0+0=0.
\end{equation}
That means $(A_iA_j^{-1},B_kB_l^{-1})=I, \forall A_i,A_j\in S(A), B_k,B_l\in S(B)$.

Thus, we can extend the two sets of operators $S(A),S(B)$ to two new sets of operators $\tilde{S}(A),\tilde{S}(B)$ as follows:
\begin{align}
& \tilde{S}(A)=\{A_{i||j}=A_iA_j^{-1}|A_i,A_j\in S(A)\}, \\
& \tilde{S}(B)=\{B_{i||j}=B_iB_j^{-1}|B_i,B_j\in S(B)\},
\end{align}
According to Eq.(\ref{eqn3}), if we select one element from each of the two new sets $\tilde{S}(A),\tilde{S}(B)$, such as $A_{i||j},B_{k||l}$ (where $i,j\in\{1,\cdots,n_A\}$, $k,l\in\{1,\cdots,n_B\}$), then the two elements satisfy the relation $(A_{i||j},B_{k||l})=I$. Thus, the new sets extended from $S(A),S(B)$ can also be used in QNK protocol.

When $n_A=n_B=2$, the extended sets of operators are $\tilde{S}(A)=\{A_1A_2^{-1}, \\A_2A_1^{-1},A_1A_1^{-1}=I,A_1A_1^{-1}=I\}$ and $\tilde{S}(B)=\{B_1B_2^{-1},B_2B_1^{-1},B_1B_1^{-1}=I,B_1B_1^{-1}=I\}$. Because the identity operator $I$ in the two sets is meaningless, the number of the operators in the two extended sets does not increase through the above method (The number of the operators in each set remains 2). Thus the extension is meaningless.
However, the extension is meaningful when $n_A,n_B\geq 3$.

\section{Discussions and Conclusions}
We study the theory of quantum no-key protocols. Firstly, a framework of quantum no-key protocols is presented. The framework is expressed in two forms: trace-preserving quantum operators and natural presentations. Secondly, we defined the information-theoretical security of quantum no-key protocol, and the security of authentication keys. Finally, two kinds of quantum no-key protocols are presented. In the first scheme, the quantum operators used by Alice and Bob are all unitary. In the second scheme, Alice and Bob use trace-preserving quantum operators, and the two sets of operators used by Alice and Bob are constructed based on two multiplicative commutative sets.

In the paper, there exists the following questions.
\begin{enumerate}
\item{}
$k,i$ are secret identification key. How to use them to identify the personal identification of Alice, Bob, or attackers. How to identify in each of the three transformations?
\item{}
How to construct operator $N_k$ and a set $I(k)$ from a given number $k$?
\item{}
How to construct a set $L(k,i)$ and a set of operators $S_k(i)$ from the given $k$ and $i\in I(k)$?
\item{}
In the quantum no-key protocols, the numbers $l_1,l_2\in L(k,i)$ are selected randomly. Whether the random selection of $l_1,l_2$ can prevent from the leakage of identification keys $k\in\mathcal{K}$ and $i\in I(k)$ during the communication?
\end{enumerate}

\section*{Acknowledgements}
This work was supported by the National Natural Science Foundation of China (Grant No. 61173157).

\section*{Appendix}
{\bf Proposition 1:} Suppose two sets of operators $S(A),S(B)$ satisfy the condition Eq.(\ref{eqn2}), then the following relations hold: $\forall i,j$
\begin{eqnarray}
&& (A_i^{-1},B_jB_i^{-1})=I,\\
&& (A_j^{-1},B_jB_i^{-1})=I,\\
&& (B_i,A_jA_i^{-1})=I,\\
&& (B_j,A_jA_i^{-1})=I.
\end{eqnarray}
{\bf Proof:} According to the condition Eq.(\ref{eqn2}), $\forall i\neq j$
\begin{eqnarray*}
&& (A_i,B_i)=N,\\
&& (A_i,B_j)=N,\\
&& (A_j,B_i)=N,\\
&& (A_j,B_j)=N.
\end{eqnarray*}
From the identity $(A_i,B_i)=N=(A_i,B_j)$, it can be deduced $B_i^{-1}A_i^{-1}B_iA_i=B_j^{-1}A_i^{-1}B_jA_i$, then
$$B_i^{-1}A_i^{-1}B_i=B_j^{-1}A_i^{-1}B_j.$$
Thus, it can be known that $(B_jB_i^{-1})A_i^{-1}=A_i^{-1}(B_jB_i^{-1})$, that is $(A_i^{-1},B_jB_i^{-1})=I$.

The other three relations can be obtained in the same way.$\hfill{}\Box$

{\bf Proposition 2:} Suppose two sets of operators $S(A),S(B)$ satisfy the condition Eq.(\ref{eqn2}),  then the following relations hold: $\forall i,j,k,l$
\begin{equation}
(A_iA_j,B_kB_l)=(A_j^2,B_l^2).
\end{equation}
{\bf Proof:} In the deduction of this proof, the four identities such as $B_lA_i=A_iB_lN$ (that is $(A_i,B_l)=N$), $A_i^{-1}B_kA_i=B_kN$ (that is $(A_i,B_k)=N$), $B_k^{-1}A_j^{-1}B_k=B_l^{-1}A_j^{-1}B_l$, and $B_lN=A_j^{-1}B_lA_j$ (that is $(A_j,B_l)=N$) are used in turn. The deduction is as follows.
$\forall i,j,k,l$,
\begin{eqnarray*}
(A_iA_j,B_kB_l)&=& (B_l^{-1}B_k^{-1})(A_j^{-1}A_i^{-1})B_kB_lA_iA_j \\
&=& (B_l^{-1}B_k^{-1})A_j^{-1}A_i^{-1}B_kA_iB_lNA_j \\
&=& B_l^{-1}B_k^{-1}A_j^{-1}B_kNB_lNA_j \\
&=& B_l^{-1}B_l^{-1}A_j^{-1}B_lNB_lNA_j \\
&=& (B_l^{-1})^2 A_j^{-1}A_j^{-1}B_lA_jA_j^{-1}B_lA_jA_j \\
&=& (B_l^{-1})^2 (A_j^{-1})^2B_l^2A_j^2 \\
&=& (A_j^2,B_l^2).
\end{eqnarray*}
$\hfill{}\Box$

{\bf Proposition 3:} $(A_{j_1}A_{j_2}A_{j_3},B_{i_1}B_{i_2}B_{i_3})=(A_{j_3}A_{j_2}A_{j_3},B_{i_3}B_{i_2}B_{i_3})$, $\forall i_1,i_2,i_3$, $\forall j_1,j_2,j_3$.

{\bf Proof:} It can be inferred from the formula $A_i^{-1}B_kA_i=A_j^{-1}B_kA_j$ that
\begin{eqnarray*}
A_i^{-1}B_{k_1}B_{k_2}B_{k_3}A_i &=& A_i^{-1}B_{k_1}A_iA_i^{-1}B_{k_2}A_iA_i^{-1}B_{k_3}A_i \\
&=& A_j^{-1}B_{k_1}A_jA_j^{-1}B_{k_2}A_jA_j^{-1}B_{k_3}A_j \\
&=& A_j^{-1}B_{k_1}B_{k_2}B_{k_3}A_j.
\end{eqnarray*}
Similarly, it can be inferred
\begin{eqnarray}
B_i^{-1}A_{k_1}^{-1}A_{k_2}^{-1}A_{k_3}^{-1}B_i &=& B_j^{-1}A_{k_1}^{-1}A_{k_2}^{-1}A_{k_3}^{-1}B_j.
\end{eqnarray}
Thus, the following result holds
\begin{eqnarray*}
(A_{j_1}A_{j_2}A_{j_3},B_{i_1}B_{i_2}B_{i_3})&=& B_{i_3}^{-1}B_{i_2}^{-1}B_{i_1}^{-1}A_{j_3}^{-1}A_{j_2}^{-1}A_{j_1}^{-1}B_{i_1}B_{i_2}B_{i_3}A_{j_1}A_{j_2}A_{j_3} \\
&=& B_{i_3}^{-1}B_{i_2}^{-1}B_{i_1}^{-1}A_{j_3}^{-1}A_{j_2}^{-1}A_{j_2}^{-1}B_{i_1}B_{i_2}B_{i_3}A_{j_2}A_{j_2}A_{j_3} \\
&=& B_{i_3}^{-1}B_{i_2}^{-1}B_{i_2}^{-1}A_{j_3}^{-1}A_{j_2}^{-1}A_{j_2}^{-1}B_{i_2}B_{i_2}B_{i_3}A_{j_2}A_{j_2}A_{j_3} \\
&=& B_{i_3}^{-1}B_{i_2}^{-1}B_{i_2}^{-1}A_{j_3}^{-1}A_{j_2}^{-1}A_{j_3}^{-1}B_{i_2}B_{i_2}B_{i_3}A_{j_3}A_{j_2}A_{j_3} \\
&=& B_{i_3}^{-1}B_{i_2}^{-1}B_{i_3}^{-1}A_{j_3}^{-1}A_{j_2}^{-1}A_{j_3}^{-1}B_{i_3}B_{i_2}B_{i_3}A_{j_3}A_{j_2}A_{j_3} \\
&=& (A_{j_3}A_{j_2}A_{j_3},B_{i_3}B_{i_2}B_{i_3}).
\end{eqnarray*}
$\hfill{}\Box$

\end{document}